\def\revtex{2}

\ifnum\revtex>0 
\documentstyle[aps,prb,preprint,epsfig,amsmath,amssymb,verbatim]{revtex}

\else
\documentclass[10pt]{article}
\usepackage{epsfig}
\usepackage[]{graphicx}
\DeclareGraphicsExtensions{.eps}
\usepackage{times}
\usepackage{a4wide}
\usepackage{amsmath,amssymb}
\usepackage{ltxtable}
\usepackage{setspace}
\usepackage{verbatim}
\usepackage{cite}

\fi

\newcommand{\lra}[1]{\langle #1 \rangle }
\newcommand{\mb}[1]{\mathbf{#1}}
\newcommand{\bds}[1]{\boldsymbol{#1}}

\begin{document}

\title{PDF model based on  Langevin equation for polydispersed
two-phase flows applied to a bluff-body gas-solid flow}
\author{Jean-Pierre Minier$^{1}$, Eric Peirano$^{2}$ and Sergio
  Chibbaro$^{3}$ \vspace{2mm} \\ 
$^{1}$Electricit\'e de France, Div. R\&D, MFTT,
6 Quai Watier, 78400 Chatou, France \\
E-mail: Jean-Pierre.Minier@edf.fr\vspace{2mm} \\
$^{2}$ADEME-DER, 500 
Route des Lucioles 06560 Valbonne France \\
E-mail: eric.peirano@ademe.fr \vspace{2mm} \\
$^{3}$Dipartimento di fisica and INFN,  Universit\`a di Cagliari, 
Cittadella Universitaria Monserrato CA, Italy \\
E-mail: chibbaro@chi80bk.der.edf.fr}
\maketitle

\begin{abstract}
The aim of the paper is to discuss the main characteristics of a
complete theoretical and numerical model for turbulent polydispersed
two-phase flows, pointing out some specific issues.
The theoretical details of the model have already been presented
[Minier and Peirano, Physics Reports, Vol. 352/1-3, 2001 ].
Consequently, the present work is mainly focused on complementary aspects,
that are often overlooked and that require particular attention.
In particular, the following points are analysed : the necessity to
add an extra term in the equation for the velocity of the fluid seen
in the case of two-way coupling, the theoretical and numerical
evaluations of particle averages and the fulfilment of the particle
mass-continuity constraint.
The theoretical model is developed within the PDF formalism.
The important-physical choice of the state vector variables is first
discussed and  the model is then expressed as a stochastic differential
equation (SDE) written in continuous time (Langevin equations) for the
velocity of the fluid seen. The interests and limitations of Langevin
equations, compared to the single-phase case, are reviewed.
From the numerical point of view, the model corresponds to an hybrid
Eulerian/Lagrangian approach where the fluid and particle phases are
simulated by different methods.
Important aspects of the Monte Carlo particle/mesh numerical method
are emphasised. 
Finally, the complete model is validated and its performance
is assessed by simulating a bluff-body case with an important
recirculation zone and in which two-way coupling is noticeable.
\end{abstract}

\section{Introduction} \label{sec:intro}
Dispersed two-phase flows, where a continuous phase (a gas or a liquid)
carries discrete particles (solid particles, droplets, bubbles, \dots),
are of great interest in environmental studies and engineering applications,
such as  dispersion of small particles in the atmosphere
or  combustion of fuel droplets in a car engine. 

To simulate these flows, the basic equations must be written: 
the Navier-Stokes equations for the fluid phase and the momentum
equation for a single particle embedded in a turbulent flow, the
latter issue still being a subject of current research. 
For small particle-based Reynolds numbers $Re_p$ (whose definition is
specified below) and particle diameters that are of the same order of
magnitude as the  Kolmogorov length scale, a general form of the
particle momentum equation has been proposed ~\cite{Gat_83,Max_83}.

In the present work, only heavy particles ($\rho_p \gg \rho_f$) are
under consideration and the equations of motion for a particle can be
written:
\begin{subequations} 
\label{exact particle eqns}
\begin{align} 
& \frac{d{\bf x}_p}{dt} = {\bf U}_p, \\
& \frac{d{\bf U}_p}{dt} = \frac{1}{\tau_{p}}({\bf U}_{s}-{\bf U}_{p})
\;  + {\bf g},
\end{align}
\end{subequations}
where ${\bf U}_{s}={\bf U}({\bf x}_{p}(t),t)$ is the fluid velocity seen,
\textit{i.e.} the fluid velocity sampled along the particle trajectory
${\bf x}_{p}(t)$, not to be confused with the fluid velocity ${\bf U}_{f}=
{\bf U}({\bf x}_{f}(t),t)$ denoted with the subscript $f$. The particle
relaxation time is defined as
\begin{equation} \label{definition taup}
\tau_{p}=\frac{\rho_p}{\rho_f}\frac{4 d_p}{3 C_D |{\bf U_r}|},
\end{equation}
where the local instantaneous relative velocity is
${\bf U_r}={\bf U}_p-{\bf U}_s$ and the drag coefficient $C_D$ is a
non-linear function of the particle-based Reynolds number, $Re_p=d_p
|{\bf U_r}|/\nu_f$, which means that $C_D$ is a complicated function of
$d_p$, the particle diameter ~\cite{Cli_78}. A very often retained
empirical form for the drag coefficient is
\begin{equation} \label{expression C_D}
C_D=
\begin{cases}
\displaystyle \frac{24}{Re_p}\left[\,1 + 0.15 Re_p^{0.687}\, \right]
 & \text{if} \; Re_p \leq 1000,  \\
0.44 & \text{if} \; Re_p \geq 1000.
\end{cases}
\end{equation}

In the present work, attention is focused on some aspects of the
problem. In particular, only  dilute incompressible gas-particle flows
are considered, so that particle-particle interactions are neglected
but two-way coupling is retained, which means that particle dispersion
and modulation of turbulence by the particles are accounted for.

The complete problem is formed by the discrete particle equations
given above, Eqs~(\ref{exact particle eqns}) to
(\ref{expression C_D}) and the field equations of the fluid phase,
the continuity and the Navier-Stokes equations, supplemented with a source 
term ${\bf S}$ that represents the force exerted by the particles on
the fluid
\begin{subequations}
\label{fluid: exact field eqs.}
\begin{align}
&\frac{\partial U_{f,j}}{\partial x_j}=0, \\
&\frac{\partial U_{f,i}}{\partial t}+U_{f,j}\frac{\partial U_{f,i}}{\partial x_j}
= -\frac{1}{\rho_f}\frac{\partial P}{\partial x_i} + 
   \nu \frac{\partial^2 U_{f,i}}{\partial x_j^2} + S_i.
\end{align}
\end{subequations}
An ``exact approach'' (in the spirit of DNS) is possible
~\cite{Boi_98}, but in practice, the exact equations of motion are not
of great help. Indeed, in the case of a large number of particles and
of turbulent flows at high Reynolds numbers, the number of degrees of
freedom is huge and one has to resort to a contracted probabilistic
description.

Following the classical approach used in single-phase turbulence, 
one can think of writing directly mean-field equations for a limited
number of particle statistics (mean velocity, kinetic energy, ...) as
in $k-\epsilon$ or $R_{ij}-\epsilon$ modelling.
This is the basis of the Eulerian approach ~\cite{Sim_93,Sim_96}.
However, due to the complex dependence of $\tau_p$ on particle
diameters and on fluid and particle instantaneous velocities, the drag term
represents a non-linear but local (in term of particle variables)
source term. The resulting closure problem that appears in the
Eulerian approach is therefore difficult.
Actually, this issue is very similar to the one appearing in the
modelling of single-phase turbulent reactive flows ~\cite{Pop_00} and,
in this case, PDF models that can treat the reactive source terms
without approximation have shown their great potential.
For the same reason, a PDF approach to polydispersed turbulent
two-phase flows is interesting. In practice, mean-field equations
($R_{ij}-\epsilon$) are used for the fluid whereas a particle pdf
equation is solved by a Monte Carlo method using a trajectory point of
view (Eulerian/Lagrangian models). The PDF model is therefore
formulated as a particle stochastic Lagrangian model (a set of SDEs).

Numerous Eulerian/Lagrangian two-phase flow models have been
proposed (most of the time with interesting and clear ideas), but
often with a discrete formulation (in time) and without making the
connection with a PDF model. When Eulerian/Eulerian (\textit{i.e.}
both phases are described with mean-field equations) and
Eulerian/Lagrangian models are compared directly, the PDF framework is
helpful to reveal that these methods do not contain the same level of
information: Lagrangian models are PDF models from which Eulerian
models can be extracted in a consistent way ~\cite{Sim_96,Pei_02}. 
The specificity of the present work is to present a Lagrangian model,
based on a Langevin equation for the velocity of the fluid seen, in
the PDF context.
The theoretical formulation of the Langevin PDF model 
has already been developed ~\cite{Pei_02,Min_01} and the purpose of 
the present paper is to give an overview of the complete theoretical 
and numerical issues insisting on complementary points. 
In the same PDF framework, an alternative formulation consists
in writing a PDF equation in analogy 
with Boltzmann kinetic equation ~\cite{Mash_03}, that is only
for particle location and velocity, without considering directly
the velocity of the fluid seen. For a comprehensive review of 
general results and methods in particle dispersion, we refer 
also to Stock's paper ~\cite{Stock_96}.

More specifically, the aims of the present paper are:
\begin{enumerate}
\item[(i)] to outline the main aspects of a PDF model, the interests
  and limitations of current state-of-the-art Langevin models and the
  keys points of numerical algorithms,
\item[(ii)] to point-out and to address new specific issues for
  Lagrangian models, such as the addition of extra terms for two-way
  coupling, numerical averages and the mean-continuity constraint,
\item[(iii)] to validate the complete model and to show how it
  performs by comparing the numerical results with experimental ones
  in a practical case.
\end{enumerate}

The paper is organised as follows: in Section \ref{sec:model}, several
mathematical notions related to stochastic modelling are clarified,
that is the equivalence between the trajectory and pdf points of view,
and the modelling strategy which is adopted in the present work (the
particle-tracking approach). The dimension of the system, that is the
dimension of the state vector, is also given based on physical
principals. In Section \ref{sec:disp}, closure proposals are put forward 
for the fluid velocity seen, in the form of Langevin equations.
Emphasis is put on the terms to be added in order to model cases with
two-way coupling. 
In Section \ref{sec:numerics}, the numerical approach is
presented. The main steps of the particle-mesh algorithm are explained,
while particular attention is devoted to the problems of defining
averages in two-phase flows and of verifying the particle
mean-continuity constraint. These models are validated in the
simulation of a practical case of gas-solid flows, Section
\ref{sec:application}.

\section{Stochastic modelling} \label{sec:model}
\subsection{Mathematical background} \label{sec:pdf_approach}
In this section, basic results, concerning the mathematical background
of the approach and the correspondence between SDEs and Fokker-Planck
equations, are recalled ~\cite{Pop_00,Gar_90,Bale_97}.
If one considers a system of N particles interacting through forces
that can be expressed as functions of variables attached to each particle
(for example, position, velocity, ...), then all available information
is contained in the state vector, ${\bf Z}$, of the complete system
\begin{equation}
{\bf Z}=( Z_1^1, Z_2^1, \ldots, Z_p^1\,;\, 
Z_1^2, Z_2^2, \ldots, Z_p^2\,;\, \ldots\,;\, Z_1^N, Z_2^N, \ldots, Z_p^N),
\end{equation}
where $Z_j^i$ represents the j-variable
attached to the particle labelled $i$. The dimension of the state vector 
is then $d=N \times p$ where $N$ is the total number of particles
and $p$ the number of variables attached to each particle. The
complete system, that is the N-particles, is closed. In classical
mechanics, the time evolution of such systems is often
described by a set of ordinary differential equations
\begin{equation}
\frac{d{\bf Z}}{dt} = {\bf A}(t,{\bf Z}),
\label{eq:closed_a}
\end{equation} 
which corresponds, in sample space, to the Liouville equation ~\cite{Gar_90}
\begin{equation}
\frac{\partial{p(t,{\bf z})}}{\partial{t}} + 
\frac{\partial}{\partial{{\bf z}}}
({\bf A}(t,{\bf z})\; p(t,{\bf z}))=0,
\label{eq:closed}
\end{equation} 

for the associated pdf, $p(t,{\bf z})$. This formulation is similar to
a pure convection problem (first-order partial derivatives in sample space).

As mentioned in the previous section, most of the time, the number of
degrees of freedom (the dimension of the state vector) is huge and
one has to resort to a reduced (or contracted) description ~\cite{Bale_97}.
Consequently,  one-particle pdf, $p(t,{\bf z^i})$ (for a particle
$i$) or  two-particle pdf, $p(t,{\bf z^i},{\bf z^j})$, etc, can be
considered instead of choosing the N-particle pdf,
$p(t,{\bf z})$. Let ${\bf Z^R}$ be a reduced state vector
(typically a one-particle state vector ${\bf Z^R}=(Z_1, Z_2, \ldots,
Z_p)$ corresponding to the $p$ variables attached to each particle).
Then, the time evolution equations, in physical space, for this
sub-system have the form:
\begin{equation}
\frac{d {\bf Z^R}}{dt} = {\bf A}(t,{\bf Z^R},{\bf Y})),
\label{eq:reduced_a}
\end{equation} 
where there is a dependence on the external variable ${\bf Y}$
(related to the particles not contained in ${\bf Z^R}$). In sample
space, the marginal pdf $p^r(t,{\bf z}^r)$ verifies
\begin{equation}
\frac{\partial{p^{r}(t,{\bf z}^r)}}{\partial{t}} 
+ \frac{\partial}{\partial{{\bf z}^r}}
\left[ \langle {\bf A}\,|\,{\bf z}^r\, \rangle \, 
p^{r}(t,{\bf z}^r)\right] =0, 
\label{eq:reduced}
\end{equation}
where the conditional expectation is defined by
\begin{equation}
\langle {\bf A}\, |\, {\bf z}^r \,\rangle = 
\int {\bf A}(t,{\bf z}^r,{\bf y})\,p({\bf y}\,|\, t,{\bf z}^r) 
d{\bf y} = \frac{1}{p(t,{\bf z}^r)}
\int {\bf A}(t,{\bf z}^r,{\bf y})\,
p(t,{\bf z}^r,{\bf y}) d{\bf y}.
\end{equation}
Eq. (\ref{eq:reduced}) is now unclosed, showing that a reduced
description  of a system implies a loss of information and the
necessity to introduce a model.

In the present paper, and for reasons presented in the next section,
the reduced system will be modelled by stochastic diffusion processes
~\cite{Pop_00,Min_01,Gar_90,Arn_74}. For such stochastic processes, the
time-evolution equations for the trajectories of the process are SDEs
written as:
\begin{equation} 
dZ^R_i= A_i(t,{\bf Z}^R(t))\, dt + B_{ij}(t,{\bf Z}^R(t)) dW_j,
\label{sde}
\end{equation}
where $W_j = (W_1,\ldots, W_n)$ is a set of independent Wiener processes
~\cite{Arn_74} and $n$ is the dimension of the reduced state vector.
In Eq. (\ref{sde}), ${\bf A} = (A_i)$ is called the drift vector and $
{\bf B} = (B_{ij})$ the diffusion matrix. SDEs require a strict
mathematical definition of the stochastic integral
~\cite{Gar_90,Arn_74} which is defined here in the {\it It\^o sense}
(these equations are referred to as Langevin equations in the physical
literature). The corresponding equation in sample space for
$p^r(t,{\bf z}^r)$ is the Fokker-Planck equation
\begin{equation}
\label{fokker-planck}
\frac{\partial p^r}{\partial t} =
-\frac{\partial}{\partial z_i^r}[\,A_i(t,{\bf z}^r)\, p^r\,] +
\frac{1}{2} \frac{\partial^2}{\partial z_i^r\partial z_j^r}
               [\,D_{ij}(t,{\bf z}^r)\, p^r\,].
\end{equation}
where $D_{ij} = B_{il}B_{lj} = (BB^*)_{ij}$ is a positive-definite
matrix. In a weak sense (when one is only interested in statistics of
the process), one can speak of an equivalence between SDEs and
Fokker-Planck equations. 

\subsection{Dimension of the state vector } \label{sec:dim}
The dimension of the reduced state vector, ${\bf Z}^R$, that is the
number of particles $N$ and the number of attached variables $p$ for
each particle, have to be determined (hereafter the upper-scripts $r$
and $R$ are dropped for the sake of simplicity). The first choice for
$N$ is done in line with current state-of-the-art models for
single-phase flows ~\cite{Pop_00}. Indeed, when the particle relaxation
time $\tau_p$ is small, particles behave as fluid particles. In
single-phase turbulence ~\cite{Pop_94}, only one-particle PDF models
are sufficiently general to be applicable to complex flows. For this
reason, our first choice is to retain a one-particle pdf description
for the particle phase in the two-phase flows under consideration
here ($N=1$).

The second choice is to select the specific variables attached to the
solid particles. Again, a closer look at single-phase PDF models
~\cite{Pop_00} might be helpful.
In single-phase flows at high Reynolds numbers, Kolmogorov theory
~\cite{Mon_75} tells us that, for a reference time scale $dt$ in the
inertial range, Lagrangian increments of the fluid velocity are well
correlated whereas increments of the fluid acceleration are nearly
uncorrelated. This indicates that for $dt$ belonging to the inertial
range, the fluid velocity is a slow variable and the fluid acceleration 
is a fast variable which can be eliminated (fast variable elimination)
~\cite{Hak_89}. Therefore, the state vector should include position and
velocity, \textit{i.e.} $({\bf x}_f,{\bf U}_f)$ ($p=2$). This is the starting
point for Langevin equation models for fluid particle velocities
~\cite{Pop_94,Min_97}. The model takes the form of a diffusion process
with a drift term linear in the velocity of the fluid seen~\cite{Min_97}
\begin{eqnarray}
\label{eq:Langevin_f1}
dx_{f,i} &=& U_{f,i}\,dt, \\
\label{eq:Langevin_f}
dU_{f,i} &=& -\frac{1}{\rho_f}\frac{\partial \lra{P}}{\partial x_i}\,dt
          + G_{ij}(U_{f,i}-\lra{U_{f,i}})\,dt
          + \sqrt{C_0\lra{\epsilon}}\,dW_i, 
\end{eqnarray}
where $\lra{P}$ is the mean pressure field, $\lra{\epsilon}$ is the mean
dissipation rate and $C_0$ is a constant given by Kolmogorov
theory ($C_0 = 2.1$). $G_{ij}$ is a matrix which depends on mean
quantities,
\begin{equation}
G_{ij} = - \frac{1}{T_L}\delta_{ij} + G_{ij}^a.
\end{equation} 
where $G_{ij}^a$ is an anisotropy matrix (depending on mean
quantities) and $T_L$ stands for a timescale given by ($k$ is the
turbulent kinetic energy)
\begin{equation}
T_L= \frac{1}{ \displaystyle (\, \frac{1}{2} + \frac{3}{4}C_0\, )} 
\frac{k}{\lra{\epsilon }}.
\end{equation} 

In the two-phase flow case, a similar reasoning ~\cite{Min_01}
suggests to include the velocity of the fluid seen in the state vector
that becomes (the fluid acceleration seen is a fast variable)
\begin{equation}
{\bf Z}=({\bf x}_p,{\bf U}_p,{\bf U}_s).
\end{equation}
This is different from the choice made in analogy to Boltzmann
equation, when one considers only ${\bf Z}=({\bf x}_p,{\bf U}_p)$ as in
kinetic models ~\cite{Ree_92,Pan_03}. Yet, we are dealing with particles
being agitated by an underlying turbulent fluid and a (slow) variable
related to the fluid, namely the velocity of the fluid seen, is
explicitely kept in the state vector. With the kinetic choice, not
only the derivatives of the fluid velocity seen have to be modelled
but also the fluid velocity seen itself.

\section{Modelling turbulent dispersion} \label{sec:disp}
With the present choice of the state vector, the stochastic process
used to describe the system has been chosen, \textit{i.e.} 
${\bf Z}=({\bf x}_p,{\bf U}_p,{\bf U}_s)$.
Following the trajectory point of view mentioned in Section
\ref{sec:pdf_approach}, a time-evolution equation for ${\bf U}_s$ has
to be proposed. This equation, together with Eqs 
(\ref{exact particle eqns}), will give the complete system of SDEs for
the components of ${\bf Z}$. Contrary to most Lagrangian models, which
are often built in a discrete setting, the current model is written in
continuous time, as Eq. (\ref{sde}), in order to be consistent with
the proposed mathematical framework. 

\begin{figure}[htbp]
\begin{center}
\epsfig{file=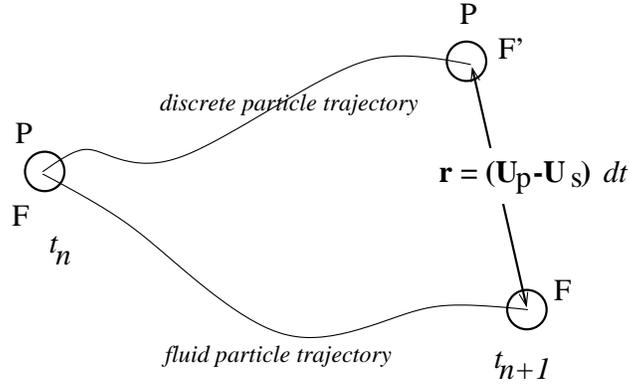,height=8cm}
\caption{Fluid element and particle paths}
\label{separation traj}
\end{center}
\end{figure}

From the physical point of view, a time-evolution equation for
${\bf U}_s$ amounts to modelling turbulent dispersion, an issue which
is more complicated than turbulent diffusion. Indeed, particle inertia
($\tau_p$) and the effect of an external force field induce a
separation of the fluid element and of the discrete particle initially
located at the same point, as represented in
Fig.~\ref{separation traj}. 
In the asymptotic limit of  small particle inertia, $\tau_p \to 0$,
and in absence of  external forces, this separation effect disappears
and  the problem of modelling diffusion is retrieved, for which the
stochastic model given by Eq.~(\ref{eq:Langevin_f}) can be
applied. For that reason, dispersion models (simulation of
${\bf U}_s$) are extensions of diffusion models (simulation of 
${\bf U}_f$).

An extensive description of the physical aspects of turbulent dispersion
has been proposed elsewhere ~\cite{Min_01,Poz_98}, so that only
the key points used to derive the stochastic model are recalled in the
next section. It is proposed to consider separately the physical
effects of particle inertia and external forces.
Two non-dimensional numbers have been introduced for that purpose:
particle inertia is measured by the Stoke number $St=\tau_p/T_L$,
and external forces by $\xi = |U_r|/u^{\prime}$, $u^{\prime}$
being a characteristic fluid turbulent velocity
($u^{\prime}=\sqrt{2k/3}$). 
The influence of these two effects on the characteristics of
${\bf U}_s$ are :
\begin{enumerate}
\item[(i)] in the absence of external forces $(\xi = 0)$, only
  particle inertia plays a role. The characteristic, or integral,
  timescale of the velocity of the fluid seen, say $T_{L}^{*}(\xi =
  0)$ is expected to vary between the fluid Lagrangian timescale,
  $T_{L}$, in the limit of low $St$ numbers, and  the Eulerian
  timescale, $T_E$, in the limit of high $St$ numbers.
\item[(ii)]Leaving out particle inertia, external forces creates mean
  drifts  $(\xi \ne 0)$ and induce a decorrelation of the velocity of
  the fluid seen with respect to the velocity of fluid particles. This
  effect is called the crossing trajectory effect (CTE) and is related
  to a mean relative velocity between particles and the fluid rather
  than an instantaneous one.
\end{enumerate}

In the model developed in the present paper, it is assumed that
$T_E$ remains of the same order of magnitude as $T_L$, which seems
actually a reasonable choice since there is little information for complex
flows. Detailed models have been proposed for the effect of particle
inertia ~\cite{Poz_98}, but in the following it will be neglected, that is
$T_{L}^{*}(\xi = 0)=T_L$. The representative picture is
now sketched in Fig. \ref{crossing} where only the mean drift induces
separation.

\subsection{Langevin equation model} \label{sec:models}
\begin{figure}[htbp]
\begin{center}
\epsfig{file=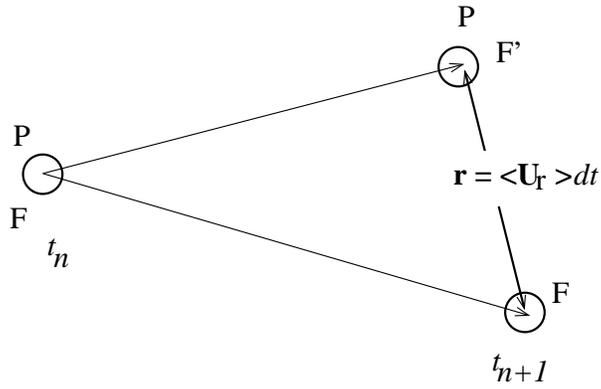,height=8cm}
\caption{Mean fluid and particle paths}
\label{crossing}
\end{center}
\end{figure}

Using the physical description of the CTE effect as due to a
mean-drift (Fig. \ref{crossing}), Kolmogorov theory can be applied, as
in the single-phase case, to suggest a dispersion model.
Indeed, let us introduce ${\bf v}(\tau,{\bf r}) = 
{\bf u}_f(t_0+\tau,{\bf x_0}+u(t_0,{\bf x_0})\tau +{\bf r})-
{\bf u}_f(t_0,{\bf x_0})$, the fluid velocity field relative to the
velocity of the fluid particle F at time $t_n$, Fig. \ref{crossing},
that is with  ${\bf u}_f(t_0,{\bf x}_0)={\bf u}_s(t_0)$, then one can
write that
\begin{equation}
d{\bf U}_s = {\bf v}(dt, \lra{ {\bf U}_r } \, dt ),
\label{disp1}
\end{equation}
where $\lra{ {\bf U}_r }= \lra{ {\bf U}_p }-\lra{ {\bf U}_s}$ is the
mean relative velocity between the discrete particle and the
surrounding fluid element. Then, the differential change, and so the
Eulerian statistics, of the  fluid velocity seen depend on the key
variables of Kolmogorov (as the fluid velocity), that is $<\epsilon>$
and $\nu$, and on the mean drift  due to the CTE effect, but not on
the instantaneous particle or fluid velocities. Since it is the mean
velocity ${\bf U}_r$ that appears in Eq. (\ref{disp1}), the Kolmogorov
theory can then be applied ~\cite{Mon_75}, to show that for
high-Reynolds number flows and for a time increment $dt$ that belongs
to the inertial range, we have
\begin{equation}
\lra{ d U_{s,i}\;dU_{s,j}}=D_{ij}(dt),
\end{equation}
where the matrix $D_{ij}$ is determined by the two scalars functions
$D_{||}$ and $D_{\bot}$ through
\begin{equation}
D_{ij} = D_{\bot}\delta_{ij} + \left[ D_{||} - D_{\bot} \right] r_ir_j,
\end{equation}
the separation vector ${\bf r}$ being in the direction of the mean
relative velocity, ${\bf r}=\lra{{\bf U}_r }/\vert\lra{{\bf U}_r}\vert $.
The functions $D_{||}$ and $D_{\bot}$ are the longitudinal and transverse
velocity correlation, respectively. Dimensional analysis yields that
in the inertial range, one can write
\begin{equation}
D_{||}(dt) = \lra{\epsilon}\, dt\,  
\alpha_{||}\left( \, \frac{ \vert \lra{{\bf U}_r}
    \vert^2}{\lra{\epsilon} dt }\, \right), 
\qquad D_{\bot}(dt) = \lra{\epsilon}\, dt\, 
\alpha_{\bot}\left( \, \frac{ \vert \lra{{\bf U}_r}
    \vert^2}{\lra{\epsilon} dt }\, \right).
\end{equation}
For the two functions $\alpha_{||}$ and $\alpha_{\bot}$, there is no
exact prediction, but in two limit cases they can be explicitely
computed. On one hand, when the mean relative velocity is small,
$\vert \lra{{\bf U}_r} \vert \ll (\lra{\epsilon} dt)^{1/2}$, for a
given time interval $dt$, the statistics of the velocity of the fluid
seen are expected to be close to the fluid ones, and thus 
$\alpha_{||} \simeq \alpha_{\bot} \simeq C_0$. 
On the other hand, when the mean relative velocity is large,
($ \vert \lra{{\bf U}_r} \vert \gg (\lra{\epsilon} dt)^{1/2}$), one can
resort to the frozen turbulence hypothesis, and in that case ($C$ is a
constant)
\begin{equation}
D_{||}(dt) \simeq C (\lra{\epsilon}\, \lra{{\bf U}_r} \, dt)^{2/3},
\qquad D_{\bot}(dt) \simeq \frac{4}{3} \,C  
(\lra{\epsilon}\, \lra{{\bf U}_r} \, dt)^{2/3},
\end{equation}
which shows that, in that limit, the two functions
$\alpha_{||}(x)$ and $\alpha_{\bot}(x)$ vary as $x^{1/3}$.
Then, the Langevin model is not supported as in the fluid case, since
it will always give a velocity correlation linear in time for each
components of $D$. Nevertheless, a useful approximation can be
proposed. Indeed, if we freeze the values of the functions
$\alpha_{||}$ and $\alpha_{\bot}$ for a certain value of the time
interval, say $\Delta t_r$, and write
\begin{equation}
D_{||}(dt) \simeq \lra{\epsilon}\, dt\,
\alpha_{||}\left( \, \frac{ \vert \lra{{\bf U}_r} \vert^2}
{\lra{\epsilon} \Delta t_r }\, \right),
\qquad D_{\bot}(dt) \simeq \lra{\epsilon}\, dt\, 
\alpha_{\bot}\left( \, \frac{ \vert \lra{{\bf U}_r} \vert^2}
{\lra{\epsilon} \Delta t_r }\, \right),
\end{equation}
a linear variation of $D_{||}(dt)$ and $D_{\bot}(dt)$, with respect to
the time interval $dt$, is now obtained. The reference time lag may be
the Lagrangian timescale which is the timescale over which fluid
velocities are correlated. And since $\lra{\epsilon} \,T_L \simeq k$,
we have
\begin{equation}  \label{structure modified}
D_{||}(dt) \simeq \lra{\epsilon}\, dt\,
\alpha_{||}\left( \, \frac{ \vert\lra{{\bf U}_r}\vert^2}{k }\,\right),
\qquad D_{\bot}(dt) \simeq \lra{\epsilon}\, dt\,
\alpha_{\bot}\left( \, \frac{\vert\lra{{\bf U}_r}\vert^2}{k }\,\right).
\end{equation}
This result suggests now a Langevin equation model which consists in
simulating ${\bf U}_s$ as a diffusion process. As explained above,
this model is only  an approximate model having less support than in
the fluid case. Indeed, the Langevin model does not yield the correct
spectrum (in the limit of large relative velocity or frozen turbulence).
However, for engineering purposes, where the macroscopic behaviour is
the real subject of interest, the important properties are the
integral time scales rather than the precise form of the spectrum.
Thus, Langevin models are ``reasonable compromises'' between
simplicity and physical accuracy at the moment. It is also clear that
much work remains to be done to improve stochastic models. 

It can be shown ~\cite{Min_01} that the general stochastic differential
equations for the fluid velocity seen process have the form (${\bf X}$
stands for fluid fields)
\begin{equation} 
dU_{s,i}= A_i(t,{\bf Z},\lra{{\bf Z}},\lra{{\bf X}} ) dt + B_{ij}(t,{\bf Z},
\lra{{\bf Z}},\lra{{\bf X}}) dW_j,
\end{equation}
where the drift vector, ${\bf A}_s$, and the diffusion matrix, 
${\bf B}_s$, have the form 
\begin{eqnarray}
\label{eq:dUs}
dU_{s,i} &=& -\frac{1}{\rho_f}\frac{\partial \lra{P} }{\partial x_i}\, dt
+ \left( \lra{U_{p,j}} - \lra{U_{f,j}} \right)
\frac{\partial \lra{U_{f,i}}}{\partial x_j}\, dt \nonumber \\
 &&-\frac{1}{T_{L,i}^*}
\left( U_{s,i}-\lra{U_{f,i}} \right)\, dt \nonumber \\
&& + \sqrt{ \lra{\epsilon}\left( C_0b_i \tilde{k}/k
+ \frac{2}{3}( b_i \tilde{k}/k -1) \right) }\, dW_i.
\end{eqnarray}
The CTE has been modelled by changing the timescales in drift and
diffusion terms  according to Csanady's analysis. Assuming, for the
sake of simplicity, that the mean drift is aligned with the first
coordinate axis (the general case is discussed elsewhere
~\cite{Min_01}), the modelled expressions for the timescales are, in
the longitudinal direction
\begin{equation}
\label{eq:CsanadyL}
T_{L,1}^{*}= \frac{T_L^*(\xi=0)} {\sqrt{ 1 + \beta^2 \displaystyle \frac{\vert
      \lra{{\bf U}_r}\vert^2}{2k/3}}},
\end{equation}
and in the transversal directions (axes labelled 2 and 3)
\begin{equation}
\label{eq:CsanadyT}
T_{L,2}^{*}= T_{L,3}^{*} = \frac{T_L^*(\xi=0)} {\sqrt{ 1 + 4\beta^2 \displaystyle
    \frac{\vert \lra{{\bf U}_r}\vert^2}{2k/3}}}~.
\end{equation}
In these equations $\beta$ is the ratio of the Lagrangian 
and the Eulerian timescales of the fluid, $\beta=T_L/T_E$,
and  $T_L^*(\xi=0)$ represents the Lagrangian time-scale 
in the absence of mean drifts but accounting for particle inertia.
As mentioned at the end of the previous section, particle inertia effect 
are neglected in the present work ~\cite{Poz_98} and we therefore assume that
 $T_{L}^{*}(\xi = 0)=T_L$.
In the diffusion matrix, a new kinetic
energy has been introduced ($b_i= T_L/T_{L,i}$)
\begin{equation}
\tilde{k}= \frac{3}{2}
\frac{\sum^3_{i=1}b_i\lra{u_{f,i}^2}}{\sum^3_{i=1}b_i}.
\end{equation}

In the absence of mean drifts, the stochastic model for ${\bf U}_s$
reverts to the Langevin equation model used in single-phase PDF
modelling ~\cite{Pop_00} and is thus free of any spurious drift by
construction. Finally, it must be emphasised that the derivation of a
satisfactory model (that is respecting a number of well-established
constraints) for particle dispersion remains an open issue. 

\subsection{Modelling two-way coupling} \label{sec:two_way_coupling}
In order to account for the influence of the particles on the fluid, a
new term is added in the momentum equation of the fluid velocity, see
Eqs (\ref{fluid: exact field eqs.}), and the fluid velocity seen,
\begin{equation}
\label{eq:model_Us}
dU_{s,i} = \left[A_{s,i}(t,{\bf Z},\lra{{\bf Z}},\lra{{\bf X}})+ \,
    A_{p \rightarrow s,i}(t,{\bf Z},\lra{{\bf Z}})\right]\, dt +
    B_{s,ij}(t,{\bf Z},\lra{{\bf Z}},\lra{{\bf X}})\; dW_j.
\end{equation}
The exact expression for this acceleration,
$A_{p \rightarrow s,i}(t,{\bf Z},\lra{{\bf Z}})$, which is induced by
the presence of the discrete particles, is not a priori known. The
underlying force corresponds to the exchange of momentum between the
fluid and the particles, but should not be confused with the total
force acting on particles since the latter includes external forces
such as gravity. The effect of particles on fluid properties is
expressed directly in the stochastic equation of ${\bf U}_s$ with a
simple stochastic model. The force exerted by one
particle on the fluid corresponds to the drag force written here as
\begin{equation}
\label{drag term}
{\bf F}_{p \rightarrow f}= - m_p \frac{{\bf U}_s-{\bf U}_p}{\tau_p},
\end{equation}
where $m_p$ is the mass of a particle. The total force acting on the
fluid element surrounding a discrete particle  is then obtained as the
sum of all elementary forces, ${\mb {F}}_{p \rightarrow f}$, and the
resulting acceleration is modelled here as ~\cite{Min_01}
\begin{equation}
\label{def: reverse force}
A_{p \rightarrow s,i} = - \frac{\alpha_p \rho_p}{\alpha_f \rho_f}
\frac{U_{p,i} - U_{s,i}}{\tau_p}.
\end{equation}

Eq.~(\ref{eq:model_Us}) is justified by the assumption that the mean
transfer rate of energy and energy dissipation $\lra{\epsilon}$ is
changed by the presence of particles, but the nature and structure of
turbulence remains the same. Therefore, Eq.~(\ref{eq:model_Us}) is
written by adding an acceleration term, 
${\bf A}_{p \rightarrow s}$, to account for
the presence of particles, while  the same closures as in the one-way
coupling case will be used for the drift vectors and the diffusion
matrices, where, once again, the mean fields $\lra{\epsilon},
\lra{U_f^2}, \dots$ are modified by the presence of the
particles. Indeed, the drift vectors and the diffusion matrices not
being affected by the nature of turbulence, remain unchanged.
In opposition to the previous hypotheses, recent results of direct
numerical simulations in the field of turbulence modulation by
particles (in isotropic  turbulence) ~\cite{Boi_98} seem to indicate
that there is a non-uniform distortion of the energy spectrum. This
could mean that, contrary to our previous assumption, the nature and
structure of the energy transfer mechanisms of turbulence  are
modified by the presence of particles. There is no precise
'geometrical' knowledge on the structure of turbulence in the presence
of discrete particles and this makes it extremely difficult to isolate
the important variables in order to modify the theory of Kolmogorov
(which is used in our closures). This problem is out of the scope of
the present paper and it remains an open question. Then, the final set
of equations for the velocity of the fluid seen are: 
\begin{eqnarray}
\label{eq:dUs_comp}
dU_{s,i} &=& -\frac{1}{\rho_f}\frac{\partial \lra{P} }{\partial x_i}\, dt
+ \left( \lra{U_{p,j}} - \lra{U_{f,j}} \right)
\frac{\partial \lra{U_{f,i}}}{\partial x_j}\, dt 
- \frac{\alpha_p \rho_p}{\alpha_f \rho_f}
\frac{U_{p,i} - U_{s,i}}{\tau_p}\,dt \nonumber \\
 &&-\frac{1}{T_{L,i}^*}
\left( U_{s,i}-\lra{U_{f,i}} \right)\, dt \nonumber \\
&& + \sqrt{ \lra{\epsilon}\left( C_0b_i \tilde{k}/k
+ \frac{2}{3}( b_i \tilde{k}/k -1) \right) }\, dW_i.
\end{eqnarray}
It is seen that the resulting Langevin equation, which is believed to
represent the simplest model for two-phase flows, contains a diagonal
but non-isotropic diffusion  matrix, $B_{s,ij}=B_{s,i}\,\delta_{ij}$.
It is also worth emphasising that the closure relations put forward
just above reflect modelling choices.  For instance, in the two-phase
flow case, the isotropic form of the diffusion matrix cannot be
obtained anymore, but it is {\em chosen} to select among different
possibilities, a diagonal diffusion matrix.

\subsection{Equivalence with the PDF approach} \label{sec:FP}
According to the arguments developed in Section \ref{sec:pdf_approach},
the complete set of SDEs (for the state vector
${\bf Z}=({\bf x}_p,{\bf U}_f,{\bf U}_s)$),
\begin{subequations} \label{eq:sde}
\begin{align}
dx_{p,i} & = U_{p,i}\, dt \\
dU_{p,i} & = \frac{1}{\tau_p}\,(U_{s,i} - U_{p,i})\, dt + g_i\, dt \\
dU_{s,i} & = \left[A_{s,i}(t,{\bf Z},\lra{{\bf Z}},\lra{{\bf X}})+ \,
  A_{p \rightarrow s ,i}(t,{\bf Z},\lra{{\bf Z}})\right]\,dt+
B_{s,ij}(t,{\bf Z},\lra{{\bf Z}},\lra{{\bf X}})\;dW_j.
\end{align}
\end{subequations}
is equivalent to a Fokker-Planck equation given in
closed form for the corresponding pdf 
$p(t;{\bf y}_p,{\bf V}_p,{\bf V}_s)$ which is, in sample space
\begin{eqnarray}
\label{eq:t_p}
&& \frac{\partial p}{\partial t} 
+  V_{p,i} \frac{\partial p}{\partial y_{p,i}} = 
- \frac{\partial}{\partial V_{p,i}}(A_{p,i}\, p\,) \notag \\
&& - \frac{\partial}{\partial V_{s,i}}
( \left[ A_{s,i}+ \lra{A_{p \rightarrow s,i}|\,
              {\bf y}_p,{\bf V}_p,{\bf V}_s}\right]\, p\,)
+ \frac{1}{2}\frac{\partial^2}{\partial V_{s,i}\partial V_{s,j}}
           ([B_s B_s^T]_{ij}\, p\,).
\end{eqnarray}
The equation for the Eulerian pdf and the resulting mean-field
equations can be found in ~\cite{Pei_02}.

\section{Numerical Issues} \label{sec:numerics}
The theoretical model developed in Section \ref{sec:disp} represents a
PDF model for the particle phase only. It does not contain any
description of the continuous phase. It is possible to extend
the PDF description to both the fluid and particle phases ~\cite{Pei_02},
which may be useful for theoretical and consistency analysis. However,
at the moment, this complete PDF approach is limited for practical
calculations, and, in the present work, a classical second-moment
approach is followed for the continuous phase. The complete numerical
model is therefore  an hybrid method and corresponds to a classical
approach referred to as Eulerian/Lagrangian in the literature, as
mentioned in Section \ref{sec:intro}. As one can see from Section
\ref{sec:disp}, the terminology is not actually adequate to 
describe the complete model (it would be better to talk of a
Moment/PDF hybrid approach), but corresponds to the numerical
approach. Indeed, from the numerical point of view, the fluid phase is
modelled by mean fields, obtained by solving partial differential
equations on a grid with an Eulerian approach, while the particle
phase is modelled by a large number of Lagrangian  particles
distributed in the domain and whose properties are obtained by solving
stochastic differential equations. It is worth emphasising that these
particles are now stochastic particles, or more precisely samples of
the underlying pdf, rather than precise models of the actual particles.
The overall numerical method is therefore an example of Monte Carlo 
particle-mesh techniques.

The numerical (particle-mesh) approach involves many issues. Some of
them have already been  treated in classical textbooks ~\cite{Hoc_88}, 
but only for deterministic equations.
The stochastic nature of the present equations brings in specific
aspects and  raises new questions. In that respect, the purpose of
this section is not to give a comprehensive description of all
issues. It is more to give an overview of the numerical method,
pointing out important  issues and, in particular, those that, in our
opinion, require additional work. More precisely, issues that have not
always been investigated or may  have been overlooked (such as
consistent discrete averages, Section \ref{sec:averages},
and mass-continuity constraint, Section \ref{continuity}), are
developed more in detail.

\subsection{General Algorithm}
\begin{figure}[htbp]
\begin{center}
\epsfig{file=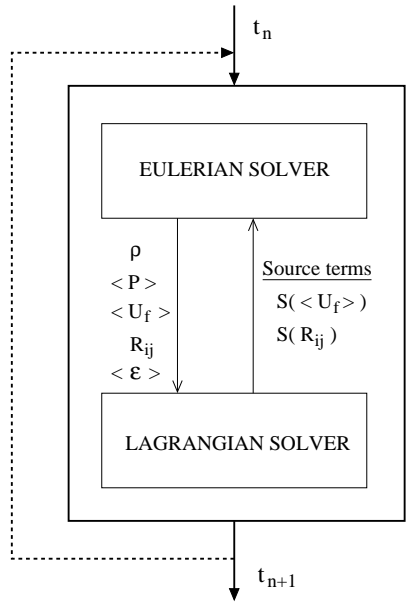,height=14cm}
\caption{Sketch of the algorithm for one time step}
\label{flow}
\end{center}
\end{figure}

The flow-chart of the code is shown in Fig. \ref{flow}. At each time
step, the fluid mean fields are first computed by solving the
corresponding partial  differential equations (RSM model) with a
classical finite volume approach. The Eulerian solver then provides
the Lagrangian solver with the fluid mean fields that are necessary to
advance particles properties. In the Lagrangian solver, the
dispersed phase is represented by a large number of particles and, as
proposed by the model, the state vector attached to each particle is
 ${\mathbf Z} = ({\bf x}_p,{\bf U}_f,{\bf U}_s)$.
Once particle properties have been updated, and in the case of two-way
coupling, where particles modify the fluid flow, source terms
accounting for momentum and energy exchange between the two phases are
also calculated and are fed back into the Eulerian solver for the next
time step computation. It is then seen that the two solvers are only
loosely-coupled. This may lead to numerical difficulties when the
particle loading is  increased, consequently the source terms become
important and the system of equations stiff. However, our present aim
is to model moderate particle loading phenomena, indeed 
particle-particle collisions have been neglected. In that range,
particles can still modify the fluid flow in a noticeable way but
source terms remain small enough so that the loosely-coupled algorithm
can still be retained.

As previously explained, the particle properties are modelled by a
vectorial SDE written as
\begin{equation} 
dZ_i= A_i(t,{\bf Z},\lra{f({\bf Z})},\lra{{\bf X}})dt + B_{ij}(t,{\bf Z},
\lra{f({\bf Z})},\lra{{\bf X}}) dW_j 
\label{sdeMK}
\end{equation}
where $f$ is a general function depending on the model and ${\bf X}$
stands for fluid fields. It is worth emphasising that the drift and
diffusion coefficients depend on statistics derived from the pdf that
is  implicitly calculated. Therefore, these SDEs are different from
standard ones ~\cite{Arn_74,Tal_95}.
Updating particles properties implies three steps: (i) projection of
$\lra{f({\bf Z})}$ and $\lra{{\bf X}}$ at particle positions, (ii)
time integration of Eq. (\ref{sdeMK}), and (iii) averaging to compute the
new values of $\lra{f({\bf Z})}$ (for stationary flows, such as the
one considered later on, ensemble averages computed in every cell are
then averaged in time, once the stationary regime has been reached.
This time-averaging procedure is very helpful to reduce statistical
noise to a negligible level ~\cite{XuPope_99,Pop_01a,Pop_01b}).
Since averaging is basically the reverse operation of projection
~\cite{Hoc_88}, these three steps correspond to two main issues:
\begin{enumerate}
\item[(i)] the first concerns the derivation of accurate numerical
  schemes for the time integration of Eq. (\ref{sdeMK}). Due to the
  non-linear nature of the equations, this is still a difficult point
  ~\cite{Tal_95,Klo_92} and, moreover, physical constraints should be
  respected ~\cite{Min_03}. This issue is briefly developed below.
\item[(ii)] The second issue is related to the exchange of information
  between the grid-based Eulerian variables, located at cell centres
  and particles which are continously distributed in the domain. At
  the moment, a NGP (nearest grid point) technique ~\cite{Hoc_88} is
  used, this represents the simplest choice but also the best one in
  terms of spatial error ~\cite{XuPope_99}. This is an important and
  attractive issue to investigate for particle-mesh methods with
  in the case of unstructured meshes and taking into account boundary
  conditions.
\end{enumerate}

\subsection{Time-integration of SDEs}
Since we are interested in the numerical approximation of statistics
derived from particles, a weak numerical scheme ~\cite{Klo_92}
(converging in law) is under consideration. A numerical scheme is said
to be of order of convergence $r$ in time, in the weak sense, if, for
any sufficiently  smooth function  
\begin{equation}
|\lra{f({\bf Z})}-\lra{f({\bf Z}^{\Delta t})}| \le C\;(\Delta t)^r,
\end{equation}
where $C$ is a constant and ${\bf Z}^{\Delta t}$ represents the
numerical approximation of ${\bf Z}$. The numerical scheme used in the
present calculations is detailed in ~\cite{Min_03}, and consequently,
in the present paper, only some points of particular importance are
emphasised. Eq. (\ref{sdeMK}) must be understood in the \textit{It\^o sense}
and it is fundamental that numerical schemes respect the It\^o
definition of the stochastic integral, in order to avoid any
inconsistency problems ~\cite{Min_03a}. The weak numerical scheme is
of order 2 in time, unconditionally stable but still explicit
~\cite{Min_03}. Another important issue is the numerical fulfilment
of physical limits ~\cite{Min_01}. Indeed, in practical engineering
calculations of complex flows, it may occur that, locally, one has
$\Delta t \gg \tau_p$, or even $\Delta t \gg T^*,\tau_p$, that is the
time-step becomes much larger than the characteristic time scales of
the system, Eqs (\ref{eq:sde}). In the first case, one should have that
${\bf U}_p \rightarrow {\bf U}_s$ and, in the second case, the model
expresses a pure diffusive behaviour in space ~\cite{Min_01,Min_03}
\begin{equation}
dx_{p,i} = \lra{x_{f,i}}\,dt + (B_{s,ij}\,T^*_L) dW_i.
\end{equation}
It is important that the numerical scheme is consistent with these
continuous limits.

\subsection{Discrete representation and numerical averages} \label{sec:averages}
Since averages are fundamental in the construction of PDF models,
it is useful to clarify the correspondence between the averages 
(defined as the mathematical expectations) and Monte Carlo estimations,
which are used in the code. In polydispersed cases, even when $\rho_p$
is constant, the mass of each particle can be different because 
of  different diameters. This suggests that even for constant density
particles, the natural definition or understanding of a mean quantity
is the \emph{mass-weighted} average. This kind of choice is somewhat
analogous to the Favre mean definition for compressible single-phase
fluid flows.

To justify this, we start by introducing a Lagrangian mass density function 
$F^L(t;{\bf y}_p,{\bf V}_p,{\bf \Psi}_p)$ where
\begin{equation}
F^L(t;{\bf y}_p,{\bf V}_p,{\bf \Psi}_p)\,d{\bf y}_p \,d{\bf V}_p
\,d{\bf \Psi}_p,
\label{mFL}
\end{equation}
is the probable mass of discrete particles in an infinitesimal volume
in sample space. As a matter of fact, attention is focused on Eulerian
averages (at a point $(t,{\bf x}_p)$ fixed in time and space), for
which the analogous Eulerian mass density function is defined by
\begin{equation}
\begin{split}
\label{eq:mcFL_p-mcFE_p}
\; F^E(t,{\bf x};{\bf V}_p,\bds{\psi}_p)
&=\; F^L(t;{\bf y}_p={\bf x},{\bf V}_p,\bds{\psi}_p) \\
&=\; \int F^L(t;{\bf y}_p,{\bf V}_p,\bds{\psi}_p)
     \,\delta({\bf x}-{\bf y}_p) \,d{\bf y}_p,
\end{split}
\end{equation}
where $F^E$ is normalised by ($\lra{\rho_p}$ is the expected density)
\begin{equation}
\label{norm}
\alpha_p(t,{\bf x})\lra{\rho_p}(t,{\bf x}) =
\int F^E(t,{\bf x};{\bf V}_p,\bds{\psi}_p)\,d{\bf
  V}_p\,d\bds{\psi}_p.
\end{equation}
$\alpha_p$ represents the probability to find particles at a given
time and position, in any state. The Eulerian mass density function
being defined, we can introduce a general average for a quantity
$H({\bf U}(t),{\bf \bds{\phi}}(t))$
\begin{equation}
\alpha_p(t,{\bf x}) \, \lra{\rho_p}(t,{\bf x}) \lra{H_p}(t,{\bf x})=
  \int H_p({\bf 
  V}_p,{\bf \bds{\Psi}_p}) {F}^E(t,{\bf x};{\bf
  V}_p,\bds{\Psi}_p)\,d{\bf V}_p\,d\bds{\psi}_p.
\end{equation}
The Lagrangian mass density function can be written from a discrete 
point of view as
\begin{equation}
F^L_N(t;{\bf y}_p,{\bf V}_p,\bds{\psi}_p)
=\sum_{i=1}^N m^i \delta({\bf y}_p-{\bf x}_p^i(t))\otimes
\delta({\bf V}_p - {\bf U}_p^i(t))\otimes \delta(\bds{\psi}_p -
\bds{\phi}_p^i(t))~
\end{equation}
where $m^i$ is the mass of the particle labelled $i$ and $N$ is the
number of samples. From Eq. (\ref{mFL}), the discrete Eulerian
mass-density functions is
\begin{equation}
F^E_N(t,{\bf x}_p;{\bf V}_p,\bds{\psi}_p)
=\frac{1}{\delta \mathcal{V}_x} \sum_{i=1}^N m^i 
\delta({\bf V}_p - {\bf U}_p^i(t))\otimes \delta(\bds{\psi}_p -
\bds{\phi}_p^i(t))~,
\end{equation}
${\delta \mathcal{V}_x}$ being a small volume around point ${\bf x}$.
Then, a numerical approximation of Eq. (\ref{norm}) is
\begin{equation}
\alpha_p(t,{\bf x}) \, \lra{\rho_p}
\simeq \frac{\sum_{i=1}^{N} m_p^i}{\delta \mathcal{V}_x},
\label{alfanum}
\end{equation}
and the numerical approximation of a particle mean quantity is
\begin{equation}
 \lra{H_p} \simeq H_{p,N} =
\frac{\sum_{i=1}^{N} m_p^i H_p({\bf U}_p^i(t),\bds{\phi}_p^i(t))}
{\sum_{i=1}^{N} m_p^i}.
\label{hnum}
\end{equation}
Convergence of the discrete approximation is ensured by the Central
Limit Theorem which shows that there exists a constant $C$ such that,
when $N \to +\infty$,
\begin{equation}
\lra{H_{p,N}}=\lra{H_p} \quad \text{and} \quad
\lra{(\lra{H_p}-H_{p,N})^2} \leq \frac{C}{N_x}.
\end{equation}
It is therefore seen that the convergence of the underlying pdf is not 
in a strong sense but in a weak sense, or to be more precise in law
~\cite{Arn_74}, since it is in fact the mean value of functions of the
stochastic process ${\bf Z}$ that converges as $N \to +\infty$,
\begin{equation}
H_{N,p}=\lra{H({\bf Z}_{N})}
\xrightarrow[N \to \infty]{}
\lra{H({\bf Z})}.
\end{equation}

\subsection{Pressure correction} \label{continuity}
It has been shown in Section \ref{sec:FP} that there is a
correspondence SDE - Fokker-Planck equation. From the pdf equation,
mean particle fields can then be extracted ~\cite{Pei_02}. In other
words, every particle stochastic model is consistent with a certain
Eulerian model ~\cite{Sim_96,Min_01} as in single-phase PDF models
~\cite{Pop_95}.

With the definition of the mean particle velocity field given in the
previous section, the corresponding particle continuity equation is
(density is constant)
\begin{equation}
\label{eq:feq_alphap_a}
\frac{\partial}{\partial t}(\alpha_p\rho_p) +
\frac{\partial}{\partial x_i}(\alpha_p\,\rho_p\lra{ U_{p,i}}) = 0.
\end{equation}
For each time step in the Lagrangian solver, the mean fields
$\alpha_p$ and $\lra{ \tilde{U}_{p,i}}$ are computed from particle
location and velocity, ${\bf x}_p$ and ${\bf U}_p$, using the numerical
approximations given in Eqs (\ref{alfanum}) and (\ref{hnum}). 
Here, we propose to modify particle velocities (not locations) so as
to enforce the mean continuity constraint, by adding a
pressure-correction field as a potential $\phi$. The corrected
particle velocity field is then
\begin{equation}
\label{eq:feq_alphap}
\lra{U_{p,i}}=\lra{\tilde{U}_{p,i}}-\frac{\partial \phi}{\partial x_i},
\end{equation}
where $\phi$ is calculated from the Poisson equation
\begin{equation}
\label{eq:feq_alphap_b}
\frac{\partial}{\partial x_i}(\alpha_p\,\rho_p\frac{\partial
  \phi}{\partial x_i})= \frac{\partial}{\partial t}(\alpha_p\rho_p) +
  \frac{\partial}{\partial x_i}(\alpha_p\,\rho_p\lra{
  \tilde{U}_{p,i}})~.
\end{equation}
The mean velocity correction term is then applied to each particle
velocity.

This pressure-correction term used here for the particle velocity is
of course similar to the classical pressure-correction step applied in
the Eulerian solver for the fluid. Yet, it is often overlooked in
Lagrangian calculations. If we consider the complete algorithm,
it is then seen that there are now two pressure-correction steps
due to the two mean-continuity equations, one for the fluid and one
for the particles. This is also a consequence of the loosely-coupled
algorithm.

\section{Numerical investigation} \label{sec:application}

\subsection{Experimental setup} \label{sec:experiment}
\begin{figure}[htbp]
\begin{center}
\epsfig{file=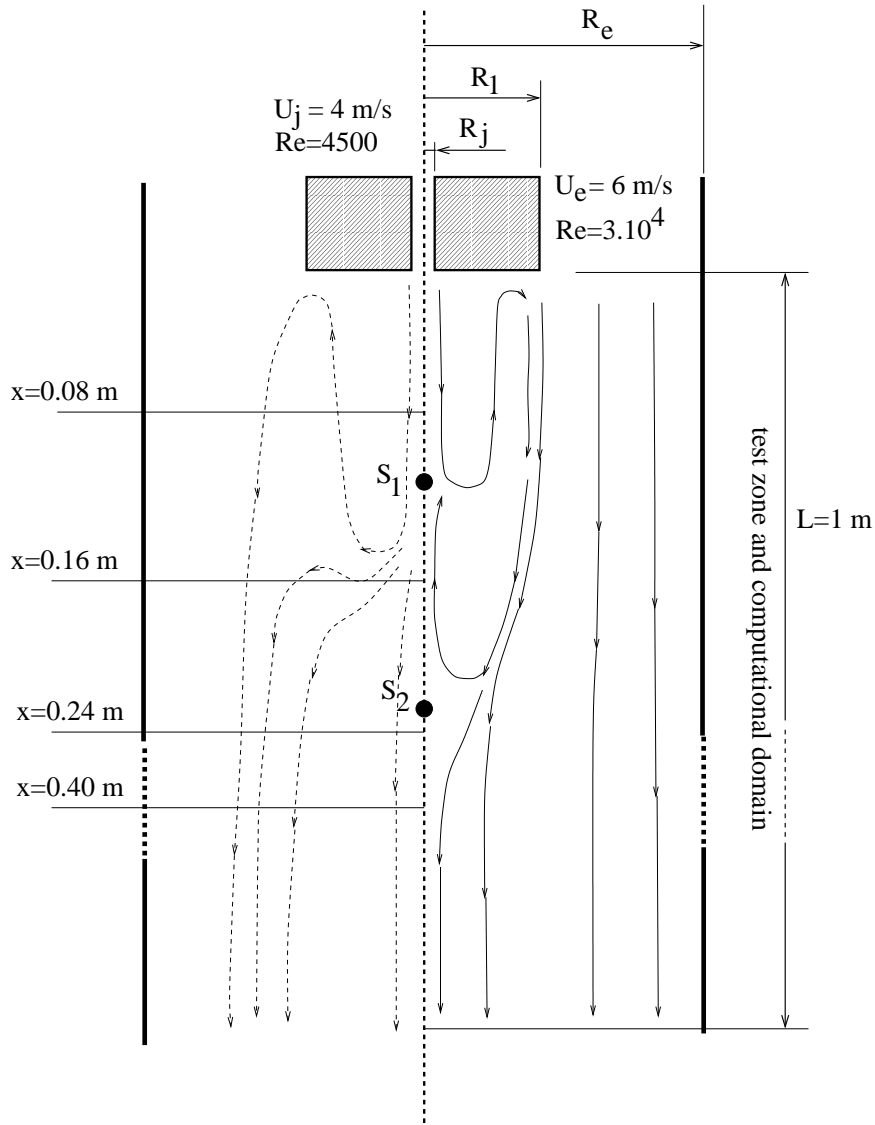,height=16cm}
\caption{ The 'Hercule' experimental setup. The mean streamlines are
  shown for the fluid (solid lines) and the particles (dashed
  lines). Two stagnation points in the fluid flow can be observed
  ($S_1$ and $S_2$). Experimental data is available for radial
  profiles of different statistical quantities at five axial distances
  downstream of the injection ($x=0.08, 0.16,0.24,0.32, 0.40 \,m$)
  (experimental data is also available on the symmetry axis). }
\label{hercule}
\end{center}
\end{figure}

The experimental setup is typical for pulverised coal combustion where
primary air and coal are injected in the centre and secondary air is
introduced on the periphery, Fig. \ref{hercule}.

This is a typical bluff-body flow where the gas (air at ambient
temperature, $T=293\,K$) is injected in the inner region and also in
the outer region where the inlet velocity is high enough to create a
recirculation zone downstream of the injection (two honeycombs were
used in the experiment in order to stabilise the flow so that no swirl
was present). Solid particles (glass particles of density
$\rho_p=2450\,kg/m^3$) are then injected from the inner cylinder with
a given mass flow rate and from there interact with the gas
turbulence. This is a coupled turbulent two-phase flow since the
particle mass loading at the inlet is high enough (22\%) for the
particles to modify the fluid mean velocities and kinetic energy. This
is also a polydispersed flow where particle diameters vary according
to a known distribution at the inlet, typically between $d_p=20 \mu m$
and $d_p=110 \mu m$ around an average of $d_p\sim 60\, \mu m$.

Experimental data are available for radial profiles (the flow is
stationary and axi-symmetric) of different statistical quantities at
five axial distances downstream of the injection ($x=0.08, 0.16,
0.24, 0.32$ and $0.40$ m). These quantities include the mean
axial and radial velocities as well as the fluctuating radial and
axial velocities for both the fluid and the particle phase. Axial
profiles along the axis of symmetry for these quantities have also
been measured. All the data was gathered using PDA measurement
techniques. Further details on the experimental setup and the
measurement techniques can be found in Ishima  \textit{et al.}
~\cite{Ish_99}.  

The 'Hercule' experimental setup is a very interesting test case for
two-phase flow modelling and numerical simulations where most of the
different aspects of two-phase flows are present. The particles are
dispersed by the turbulent flow but in return modify this
one. Furthermore, the existence of a recirculation zone where
particles interact with negative axial fluid velocities constitutes a
much more stringent test case compared to cases where the fluid and
the particle mean velocities are of the same sign (the problem is then
mostly confined to radial dispersion issues).

\subsection{Results and discussion} 
\label{sec:results}
All the results were obtained by using the ESTET 3.4 software on a
HP-C3000 workstation. In all numerical computations, the axi-symmetry
property was used: a two-dimensionnal curvilinear mesh with
$74\times 3 \times 142$ nodes was generated. The sensitivity to the
various parameters of the numerical investigation was accurately
studied. In particular, independence with respect to the time step was
checked. A uniform time step, $\Delta t=10^{-3}$~s, was then used in all
computations.

The computations were carried out with a $R_{ij}-\epsilon$ turbulence
model, which is based on the standard IPM model ~\cite{Laun_75,Laun_89}.
 Actually, this choice is satisfying from the point of view of
the consistency with the stochastic model. It is known that there is
a rigorous correspondence  between the Lagrangian stochastic models
and the second-order closures in the  case of turbulent single-phase
flows ~\cite{Pop_95}.

In the two-phase flow calculations, particles were injected when the
single-phase flow stationary regime was reached (as the limit of
the unstationary regime) before the introduction of the discrete
particles in the domain. About 1000 time steps were computed for the
single-phase problem. Around 400 to 500 additional time steps were
needed to reach the stationary regime for the two-phase flow
situation (around 14000 particles were at this stage present in the
domain). Statistics extracted from the particle data set were then
averaged in time (for about 1000 time steps) to reduce the statistical
noise.

The computational performances are shown in Table
\ref{tab:perf}. Normally, Lagrangian algorithms require much more
computational time than the Eulerian eddy-viscosity models
~\cite{Min_01}. In this case, for the same number of computational
elements (either mesh points or nodes), they appear comparable. The
computational requirements for the Eulerian solver is increased due to
the use of a full second-order turbulence model which implies the
numerical resolution of 6 coupled partial differential equations for the
fluctuating velocities (added to the 3 equations for the mean momentum)
compared to only 1 for eddy-viscosity models.

The experimental set of measures provides data both along the axis and
in cross sections at various points in the domain. The comparison is
made in all directions and at all cross sections of measures. The
cross section at $x=0.16$ is located within the recirculation zone
while the cross section at $x=0.4$ is located downstream of the limit
of the recirculation zone.

The overall agreement between experimental data and the computed
profiles is good. In particular, the particle fluctuating velocity is
well reproduced both in shape and in magnitude.

\begin{figure}[htbp]
\begin{center}
\epsfig{file=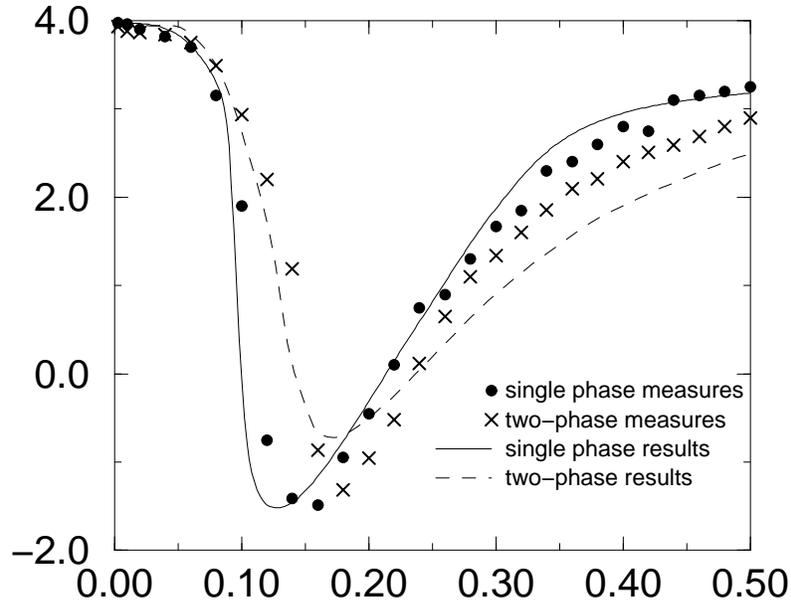,height=15cm,angle=-90}
 \caption{\label{Vit_Axe} {Mean fluid velocity along the axis in single and
two-phase flow.}}
\end{center}
\end{figure}

\begin{figure}[htbp]
\begin{center}
\epsfig{file=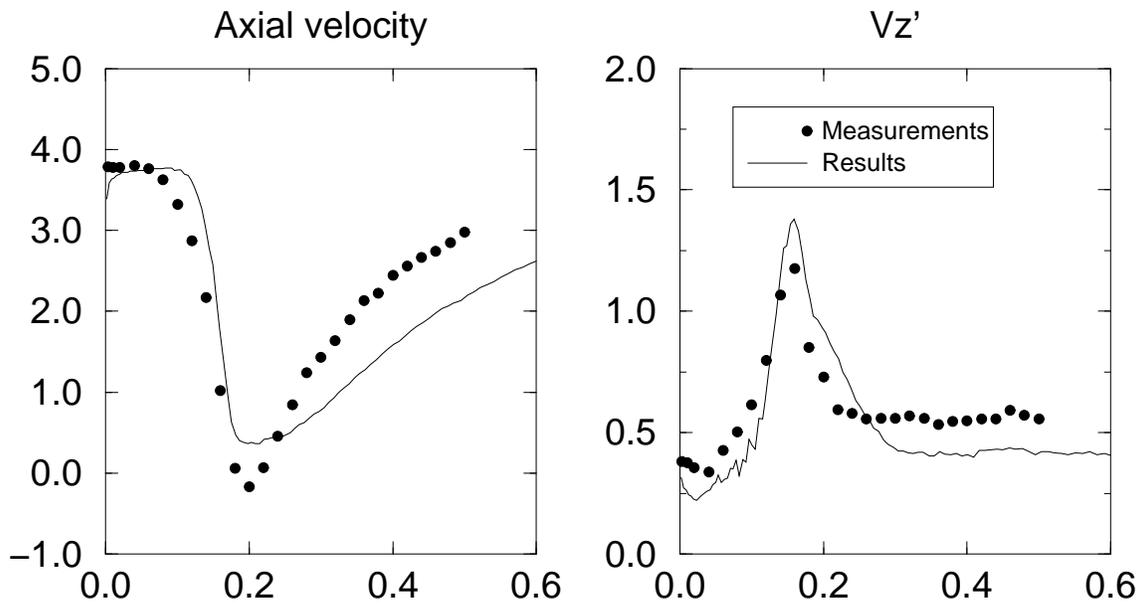,height=8cm}
\caption{Profiles of axial particle velocity along the axis (mean
  and fluctuating velocities)}
\label{VitAxe_diph}
\end{center}
\end{figure}

\begin{figure}[htbp]
\begin{center}
\epsfig{file=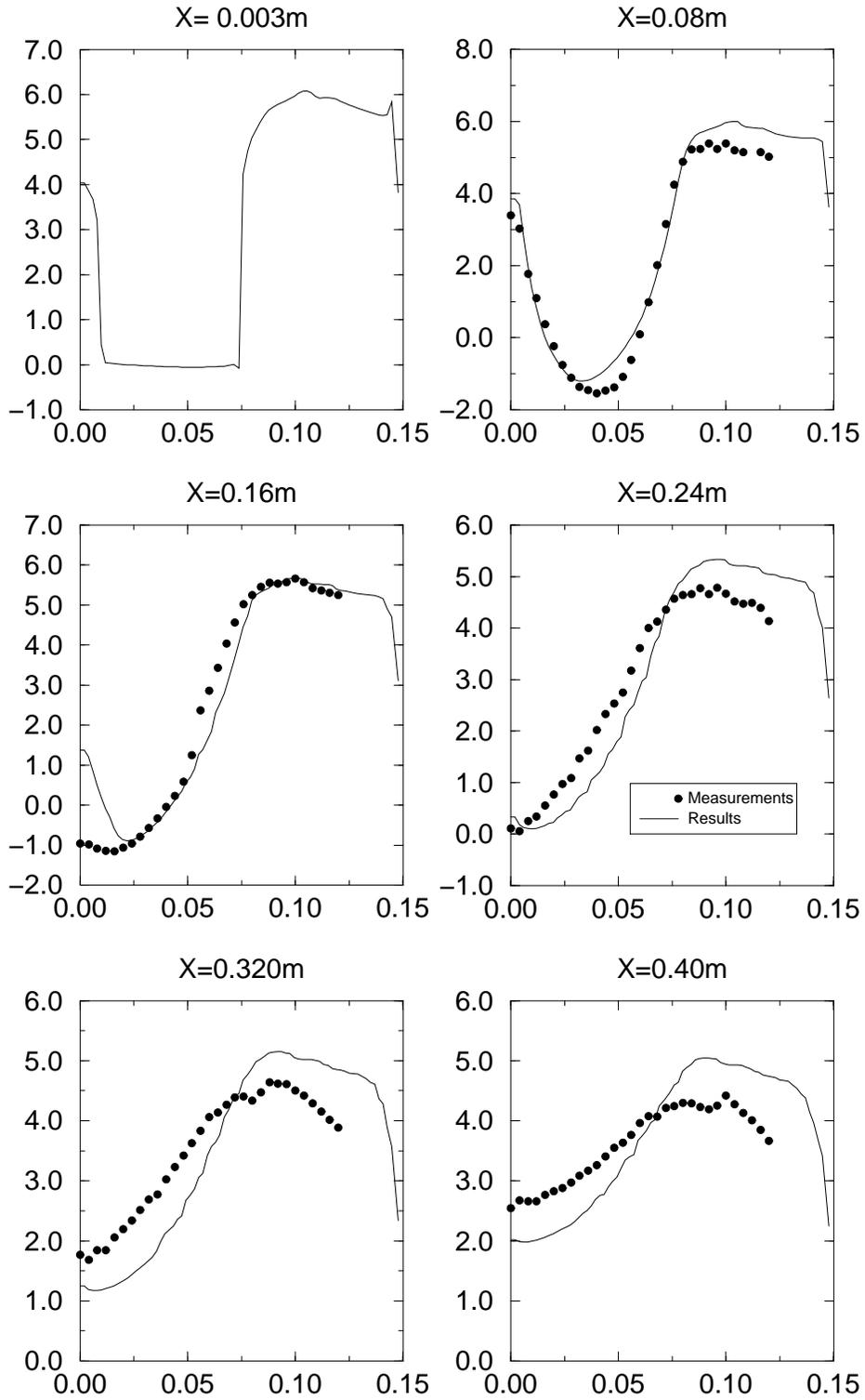,height=20cm}

\caption{Mean axial fluid velocity in two-phase simulation}
\label{Vit_Prof_fluide}
\end{center}
\end{figure}

\begin{figure}[htbp]
\begin{center}
\epsfig{file=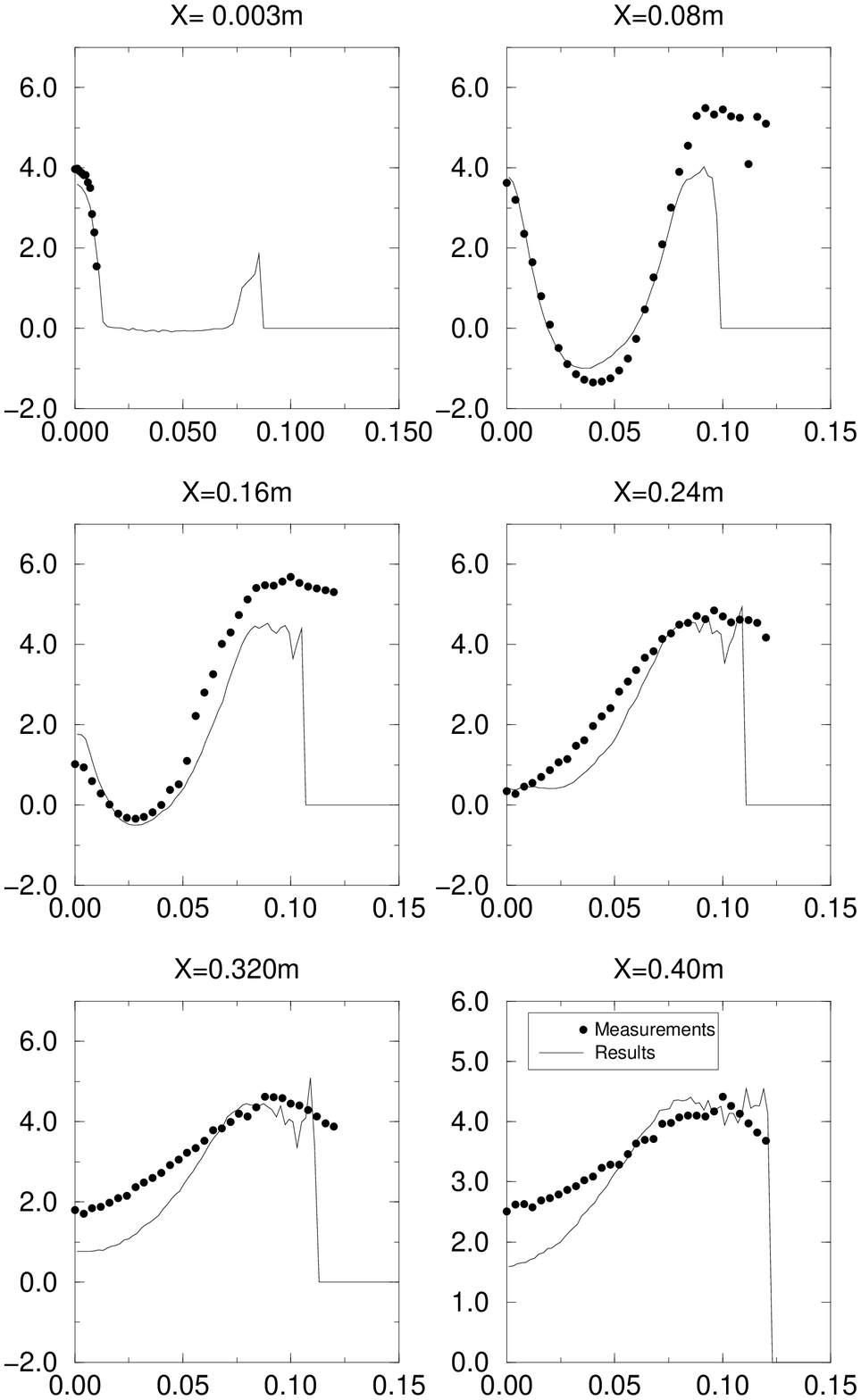,height=20cm}
\caption{Profiles of mean axial particle velocity}
\label{VitVa_Part}
\end{center}
\end{figure}

\begin{figure}[htbp]
\begin{center}
\epsfig{file=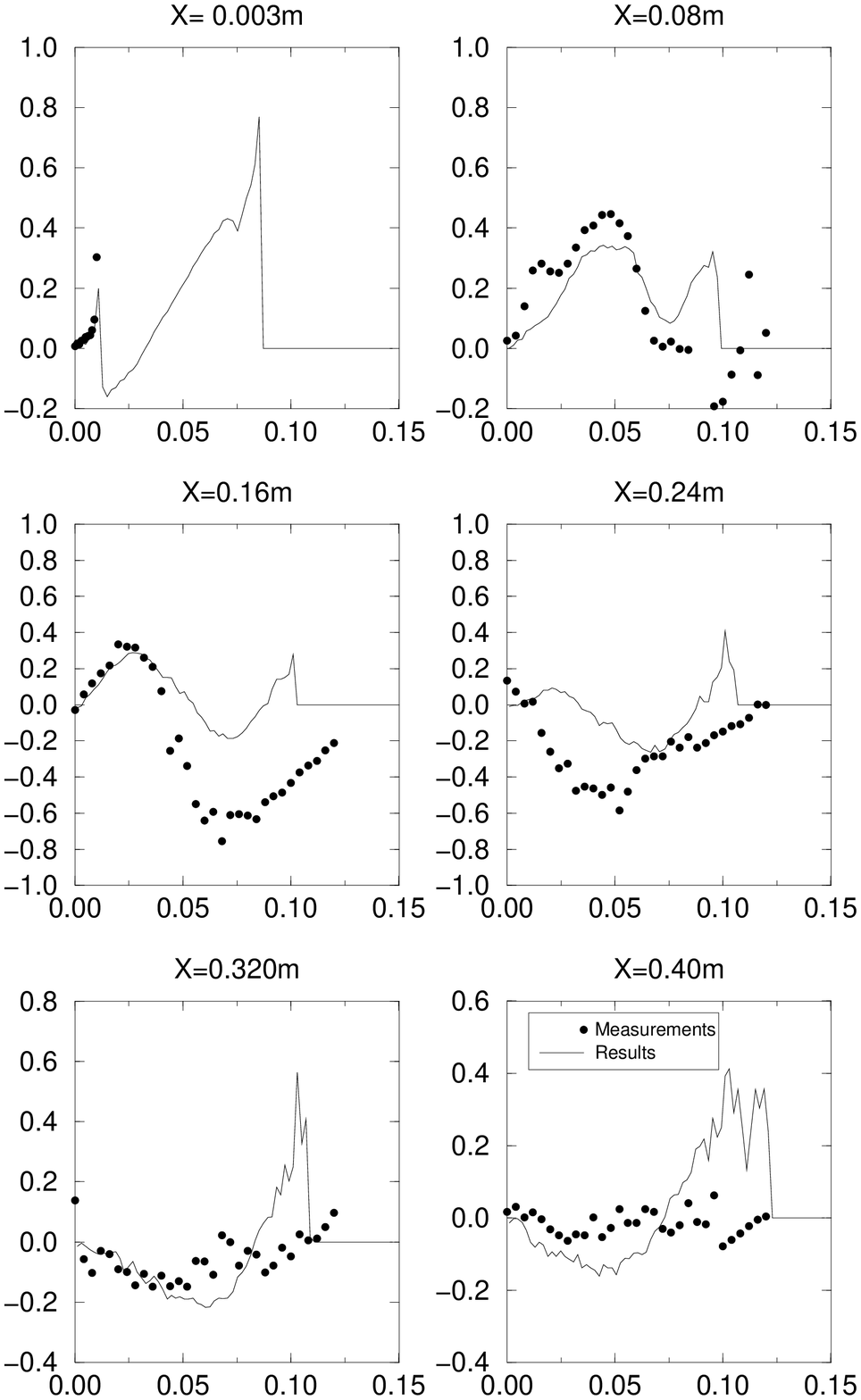,height=20cm}
\caption{Profiles of mean radial particle velocity}
\label{VitVr_Part}
\end{center}
\end{figure}

\begin{figure}[htbp]
\begin{center}
\epsfig{file=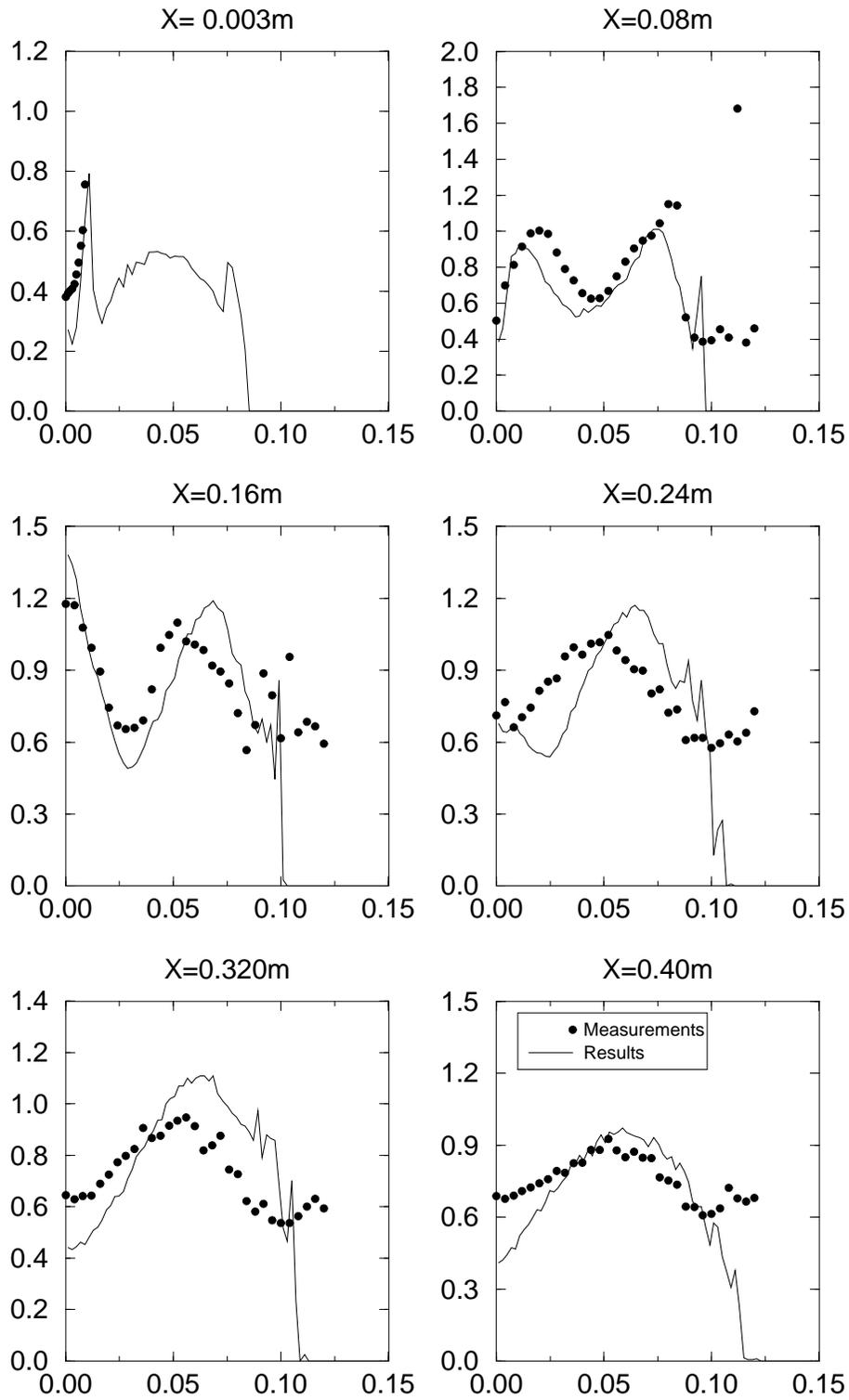,height=20cm}
\caption{Profiles of axial particle
  fluctuating velocity }
\label{VitFVa_Part}
\end{center}
\end{figure}

In Fig. \ref{Vit_Axe} and \ref{VitAxe_diph}, the mean fluid and
particle velocities along the axis are shown. It is noticeable that
the comparison between the computed results and the experimental
findings for the two-phase flow in presence of two-way coupling is
worse than in single-phase computation. The same effect characterises
both the mean fluid and the particle profiles. They results less well
reproduced in two domain zones, although the qualitatively agreement
remains good. The point of recirculation is overestimated and the
velocity slope after it is underestimated. This effects indicates the
necessity of further studies on the coupling between particles and the
fluid. It is worth noting that these effects are limited to the
behaviour along the axis. In Fig. \ref{VitVa_Part}-\ref{VitFVa_Part}
first two statistical moments of the particles velocity ($\lra{U_{p}},
\lra{W_{p}},\sqrt{\lra{u_{p}^{'2}}}$) are shown, without
smoothing. The difference between experimental data and computed
results at the axis ($x=0$) does not influence the computation in the
rest of the domain. In Fig. \ref{Vit_Prof_fluide} we show the profiles
of the fluid mean  axial velocity, where an analogous behaviour is
present, with a satisfactory agreement in the whole domain except the
values on the axis.

\section{Conclusions}
In this paper, a theoretical and numerical model for particle
turbulent polydispersed two-phase flows has been presented. 
The theoretical model is a PDF model and, in practice, appears as
a Lagrangian stochastic model. It consists in the simulation of a 
large number of stochastic particles which simulate the behaviour of 
real particles dispersed in the fluid. Each particle is defined by a set of 
variables and the selection of these state variables represents an
important choice from the physical point of view.
At present, the state variables attached to each particle include
particle position, particle velocity and the fluid velocity seen.
The present model is developed as a diffusion stochastic process
for the velocity of the fluid seen. This is similar to single-phase
turbulence, but the extension to the two-phase flow case requires
additional assumptions in the application of the Kolomogorov
hypotheses, as detailed in Section~\ref{sec:disp}. A specific point
is that, in the case of two-way coupling, an extra term is needed
in the stochastic equation for the velocity of the fluid seen in
order to be consistent with the mean field equations for the fluid phase.

From the numerical point of view, an hybrid Eulerian/Lagrangian, or
moment/Monte Carlo, approach is discussed. At each time step, the 
fluid phase is computed with an Eulerian code which provides the 
Lagrangian module with mean fluid quantities. The particles are then 
tracked and source terms representing the momentum and kinetic energy
exchanges are evaluated to be included in the Reynolds stress equations. 
This corresponds to a classical approach, but new aspects have been
emphasized. In particular, apart from considerations on numerical
schemes and the evaluation of particle means, the necessity of a 
correction to satisfy particle continuity equation has been stressed.

The interests and capabilities of the model have been illustrated
by the computation of a test case representative of an engineering
situation. Numerical predictions are in good agreement with the 
experimental ones and can be regarded as a validation of the model. 

Some of the current developments to this work aim at improving
numerical aspects (variance reduction technique for the computational
efficiency, new methods to compute statistical averages) and at
improving the physics of the model in the near-wall region (boundary
layer).

\section{Acknowledgements}
We would like to thank Dr. Mehdi Ouraou for his fruitful collaboration 
in numerical simulations.


\begin{thebibliography}{10}

\bibitem{Gat_83}
R.~Gatignol.
\newblock The {F}ax{\'e}n formulae for a rigid particle in an unsteady
  non-uniform {S}tokes flow.
\newblock {\em Journal de M{\'e}canique Th{\'e}orique et Appliqu{\'e}e},
  1(2):143--160, 1983.

\bibitem{Max_83}
Maxey M.R. and Riley J.J.
\newblock Equation of motion for a small rigid sphere in a nonuniform flow.
\newblock {\em Phys. Fluids}, 26(4):883--889, 1983.

\bibitem{Cli_78}
R.~Clift, J.~R. Grace, and M.~E. Weber.
\newblock {\em Bubbles, Drops and Particles}.
\newblock Academic Press, New York, 1978.

\bibitem{Boi_98}
M.~Boivin, O.~Simonin, and K.~D. Squires.
\newblock Direct numerical simulation of turbulence modulation by particles in
  isotropic turbulence.
\newblock {\em J. Fluid Mech.}, 375:235--263, 1998.

\bibitem{Sim_93}
O.~Simonin, E.~Deutsch, and J-P. Minier.
\newblock Eulerian prediction of the fluid/particle correlated motion in
  turbulent two-phase flows.
\newblock {\em Applied Scientific Research}, 51:275--283, 1993.

\bibitem{Sim_96}
O.~Simonin.
\newblock Continuum modelling of dispersed two-phase flows.
\newblock {\em Combustion and Turbulence in Two-Phase Flows, Lecture Series
  Programme, Von Karman Institute}, 1996.

\bibitem{Pop_00}
S.~B. Pope.
\newblock {\em Turbulent Flows}.
\newblock Cambridge University Press, 2000.

\bibitem{Pei_02}
E.~Peirano and J-P. Minier.
\newblock A probabilistic formalism and hierarchy of models for polydispersed
  turbulent two-phase flows.
\newblock {\em Phys. Rev. E}, 65(046301):1--18, 2002.

\bibitem{Min_01}
J-P. Minier and E.~Peirano.
\newblock The {PDF} approach to polydispersed turbulent two-phase flows.
\newblock {\em Physics Reports}, 352(1--3):1--214, 2001.

\bibitem{Mash_03}
F. Mashayek  and R.V.R. Pandya.
\newblock Analytical Description of Particle/Droplet-Laden Turbulent Flows.
\newblock {\em Progress in Energy and Combustion Science}, 29 (4):329-378, 2003.

\bibitem{Stock_96}
D.E. Stock.
\newblock Particle dispersion in flowing gases.
\newblock {\em J. Fluids Eng.}, 118:4-17, 1996.

\bibitem{Gar_90}
C.~W. Gardiner.
\newblock {\em Handbook of Stochastic Methods for Physics, Chemistry and the
  Natural Sciences}.
\newblock Springer-Verlag, Berlin, $2^{nd}$ edition, 1990.

\bibitem{Bale_97}
R.~Balescu.
\newblock {\em Statistical dynamics: matter out of equilibrium}.
\newblock Imperial College Press, London, 1997.

\bibitem{Arn_74}
L.~Arnold.
\newblock {\em Stochastic Differential Equations: Theory and Applications}.
\newblock Wiley, New-York, 1974.

\bibitem{Pop_94}
S.~B. Pope.
\newblock Lagrangian pdf methods for turbulent reactive flows.
\newblock {\em Ann. Rev. Fluid Mechanics}, 26:23--63, 1994.

\bibitem{Mon_75}
A.~S. Monin and A.~M. Yaglom.
\newblock {\em Statistical {F}luid {M}echanics}.
\newblock MIT Press, Cambridge, Mass, 1975.

\bibitem{Hak_89}
H.~Haken.
\newblock Synergetics: an overview.
\newblock {\em Rep. Prog. Phys.}, 52:515--533, 1989.

\bibitem{Min_97}
J-P. Minier and J.~Pozorski.
\newblock Derivation of a pdf model for turbulent flows based on principles
  from statistical physics.
\newblock {\em Phys. Fluids}, 9(6):1748--1753, 1997.

\bibitem{Ree_92}
M.~W. Reeks.
\newblock On the continuum equations for dispersed particles in nonuniform
  flows.
\newblock {\em Phys. Fluids A}, 4(6):1290--1303, 1992.

\bibitem{Pan_03}
R.V.R. Pandya and F.~Mashayek.
\newblock Non-isothermal dispersed phase of particles in turbulent flow.
\newblock {\em J. of Fluid Mech.}, 475:205--245, 2003.

\bibitem{Poz_98}
J.~Pozorski and J-P. Minier.
\newblock On the lagrangian turbulent dispersion models based on the langevin
  equation.
\newblock {\em Int. J. of Multiphase Flow}, 24:913--945, 1998.

\bibitem{Hoc_88}
R.~W. Hockney and J.~W. Eastwood.
\newblock {\em Computer simulation using particles}.
\newblock Adam Hilger, New-York, 1988.

\bibitem{XuPope_99}
J.~Xu and S.~B. Pope.
\newblock Assessment of numerical accuracy of {PDF}/{M}onte-{C}arlo methods for
  turbulent reacting flows.
\newblock {\em J. Comp. Phys.}, 152:192, 1999.

\bibitem{Pop_01a}
P.~Jenny, S.B. Pope, M.~Muradoglu, and D.A. Caughey.
\newblock A hybrid algorithm for the joint {PDF} equation of turbulent reactive
  flows.
\newblock {\em J. Comput. Phys.}, 166:218--252, 2001.

\bibitem{Pop_01b}
P.~Jenny, M.~Muradoglu, K.~Liu, S.B. Pope, and D.~A. Caughey.
\newblock Pdf simulations of a bluff-body stabilized flow.
\newblock {\em J. Comput. Phys.}, 169:1--23, 2001.

\bibitem{Tal_95}
D.~Talay.
\newblock {\em Simulation of Stochastic Differential Equation, in {\em
  Probabilistic Methods in Applied Physics}}.
\newblock Springer-Verlag, Berlin, 1995.
\newblock P. Kree and W. Wedig.

\bibitem{Klo_92}
P.E. Kloeden and E.~Platen.
\newblock {\em Numerical solution of stochastic differential equations}.
\newblock Springer-Verlag, Berlin, 1992.

\bibitem{Min_03}
J-P. Minier, E.~Peirano, and S.~Chibbaro.
\newblock Weak first- and second order numerical schemes for stochastic
  differential equations appearing in lagrangian two-phase flow modelling.
\newblock {\em Monte Carlo Meth. and Appl.}, 9(2):93--133, 2003.

\bibitem{Min_03a}
J-P. Minier, R.~Cao, and S.B. Pope.
\newblock Comment on the article "an effective particle tracing scheme on
  structured/unstructured grids in hybrid finite volume/pdf monte carlo
  methods" by li and modest.
\newblock {\em J. Comput. Phys.}, 186:356--358, 2003.

\bibitem{Pop_95}
S.B. Pope.
\newblock Particle method for turbulent flows: integration of stochastic model
  equations.
\newblock {\em Journal of Computational Physics}, 117:332--349, 1995.

\bibitem{Ish_99}
T.~Ishima, J.~Borr\'ee, P.~Fanouill\`ere, and I.~Flour.
\newblock Presentation of a data base: confined bluff body flow laden with
  solid particles.
\newblock {\em 9$^{th}$ workshop on two-phase flow predictions},
  Martin-Luther-Universit\"at, Halle-Wittenburg, Germany, April 13-16 1999.

\bibitem{Laun_75}
B.E. Launder, G.J. Reece  and W. Rodi.
\newblock Progress in the development of a Reynolds stress turbulence closure.
\newblock {\em J. Fluid Mech}, 68:537, 1975.

\bibitem{Laun_89}
B.E. Launder.
\newblock Second-moment closure: present.. and future?.
\newblock {\em Int.J.Heat Fluid Flow}, 10(4):282, 1989.

\end{thebibliography}

\newpage

\begin{table}[htb]
\begin{center}
\begin{tabular}{|l|c|c|}
& CPU time for $1000$ nodes/time step & CPU time for $1000$
particles/time step  \\
\hline
\multicolumn{1}{|l|}{ Eulerian solver}   & $0.20\;s$ &     \\
\hline
\multicolumn{1}{|l|}{ Lagrangian solver} &   & $0.17 \;s$   \\
\end{tabular}
\caption{Computational performances}
\label{tab:perf}
\end{center}
\end{table}

\newpage

\listoffigures

\newpage

\newpage

\newpage

\newpage

\newpage

\newpage

\newpage

\clearpage

\end{document}